\documentclass{osa-article}

\journal{oe}

\articletype{Research Article}
\usepackage{pifont}
\newcommand{\cmark}{\ding{51}}%
\newcommand{\xmark}{\ding{55}}%

\begin{document}

\title{Modulation leakage vulnerability in continuous-variable quantum key distribution}

\author{Nitin~Jain,\authormark{1,*} Ivan~Derkach,\authormark{2,$\ddagger$} Hou-Man~Chin,\authormark{1,3} Radim~Filip,\authormark{2} Ulrik~L. Andersen,\authormark{1} Vladyslav~C. Usenko,\authormark{2} and Tobias~Gehring\authormark{1,$\dagger$} }

\address{\authormark{1}Center for Macroscopic Quantum States (bigQ), Department of Physics, Technical University of Denmark, 2800 Lyngby, Denmark\\
\authormark{2}Department of Optics, Faculty of Science, Palacky University, 17. listopadu 50, 772 07 Olomouc, Czech Republic\\
\authormark{3}Department of Photonics, Technical University of Denmark, 2800 Lyngby, Denmark}

\email{\authormark{*}nitin.jain@fysik.dtu.dk} 
\email{\authormark{$\ddagger$}ivan.derkach@upol.cz} 
\email{\authormark{$\dagger$}tobias.gehring@fysik.dtu.dk} 



\begin{abstract}
Flaws in the process of modulation, or encoding of key bits in the quadratures of the electromagnetic light field, can make continuous-variable quantum key distribution systems susceptible to leakage of secret information. Here, we report such a modulation leakage vulnerability in a system that uses an optical in-phase and quadrature modulator to implement a single sideband encoding scheme. The leakage arises from the limited suppression of a quantum-information-carrying sideband during modulation. Based on the results from a proof-of-concept experiment, we theoretically analyse the impact of this vulnerability. Our results indicate that the leakage reduces the range over which a positive secret key can be obtained, and can even lead to a security breach if not properly taken into account. We also study the effectiveness of additional trusted noise as a countermeasure to this vulnerability.  
\end{abstract}

\section{Introduction}
Quadrature modulation played a significant role in the revival of classical optical communication and the inception of continuous-variable (CV) quantum optical communication at the turn of this century~\cite{Griffin02, Kikuchi2010, Ralph99, Grosshans02}. Information encoded in the amplitude and phase quadratures of the electric field, usually denoted by I and Q in classical communication or $x$ and $p$ in quantum communication, is decoded using coherent detection. The main difference between quantum and classical communication using optical modulation is that in the former, the signal states are typically much weaker than in the latter. Any two such quantum states are then non-orthogonal in practice, i.e., they exhibit an overlap in phase space. This property of non-orthogonality, together with the no-cloning theorem and Heisenberg's uncertainty principle, forms the bedrocks of quantum key distribution (QKD), a cryptographic method that facilitates secure communication~\cite{bennett1984, Scarani2009, Diamanti2015, Pirandola2020}.

An optical coherent state CVQKD transmitter randomly modulates the output of a coherent laser source  along the $x$ and/or $p$ quadratures. In the so-called sideband encoding approach~\cite{BachorRalph}, the information carried by the light beam leaving the transmitter can be described in the form of modulation sidebands: coherent states are generated as a result of weak modulation applied at frequency (side-)bands shifted away from the optical carrier~\cite{Lance2005}. 
After being exposed to loss and noise on the quantum channel, the sidebands are measured by the CVQKD receiver using a local oscillator (LO) assisted coherent detector to decode the information in the quadrature(s). 

These steps are performed as a part of a `CVQKD protocol' that allows the transmitter (Alice) and receiver (Bob) to share correlated bitstreams, which are used as secret keys for encryption after some classical data processing~\cite{Diamanti2015, Pirandola2020}. Security of the key is assessed by evaluating a lower bound on the final key length, which characterizes the information advantage of Alice and Bob over an eavesdropper (Eve), assumed to control the quantum channel. A non-zero key length assures Alice and Bob that Eve possesses at most an insignificant knowledge of the key, while a zero value implies the channel to be too unsafe for exchanging confidential messages. 

Realistic cryptographic systems, whether quantum or classical, are however vulnerable to side channels that lead to security loopholes in both design and implementation~\cite{Jain2016, Pirandola2020}. Such loopholes can destroy the security assurance: in QKD systems, Eve obtains significant information about the shared key without leaving any footprints. CVQKD systems too have been known to be prone to attacks due to device imperfections and operational limitations~\cite{Jouguet2012, Ma2013, Jouguet2013, Stiller2015, Qin2016, Zhao2019}. 

Here, we experimentally demonstrate and theoretically analyze a vulnerability due to modulation leakage~\cite{Derkach2017} in a CVQKD setup that implements an optical single sideband (OSSB) encoding scheme using an in-phase and quadrature (IQ) modulator. OSSB encoding is a technique where one of the two sidebands around the optical carrier is eliminated, effectively resulting in a single (modulated) sideband at the output of the transmitter. Apart from being spectrally efficient and immune to dispersion related issues~\cite{Smith1997}, OSSB modulation potentially offers better noise performance for CVQKD systems by placement of sidebands in a manner that avoids the noisy carrier during modulation and the low-frequency noise region during detection~\cite{Chin2020}. 

However, any practical IQ modulator is capable of only finite sideband suppression, so information about the random modulation along $x$ and/or $p$ at the transmitter is leaked on the quantum channel through the suppressed band. Eve may be able to access this suppressed band without alerting the legitimate parties, and can thus obtain more information about the key than estimated. Through a proof-of-principle experiment and ensuing security analysis, we show that as we (intentionally) reduce the sideband suppression by 20 dB (starting with the best possible value of $\sim 24\,$dB), the leakage of the secret key rises from 0.063 to 0.19 bits/symbol for reverse reconciliation (RR) and from 0.15 to 0.99 bits/symbol for direct reconciliation (DR) techniques. One way to address this security issue is to lower the bound on the secret key length, after having quantified the influence of the leakage on the Holevo information. We also investigate the conditions under which noise sources not controlled by Eve could reduce this penalty. 

In the last 5 years, there has been a gradual shift in CVQKD setups to replace the discrete amplitude and phase modulators with an IQ modulator to prepare phase-shift keying or Gaussian constellations~\cite{Qu2016, Kleis2017, Brunner2017, Laudenbach2019}. This move offers a compact design, potential cost benefits, and may also improve resilience to a Trojan-horse attack (through the reduction of a fiber connection) \cite{Jain2014, Stiller2015}. However, poor sideband suppression, which could arise due to sub-optimal settings of the DC bias control of the IQ modulator or due to finite manufacturing tolerance and RF mismatch, can lead to insecure keys if Alice and Bob do not take the leakage into account. Although the demonstrated vulnerability affects CVQKD systems based on single-sideband encoding only, leakage can also occur due to production of higher-order sidebands (in the so-called `strong modulation' regime), or more generally, due to multiple back-reflections inside Alice's station~\cite{Jain2014, Jouguet2012, Stiller2015}. 

The paper is organized as follows: we first develop a basic model of an IQ modulator and describe the conditions in which the leakage gets manifested. We then describe the attack model that treats the suppressed sidebands as excessive modulation. After detailing the experiment and the proof-of-concept implementation of Eve's attack strategy, we present the measurement results. Following a discussion on the impact of the attack and countermeasures, we conclude this work.  

\section{Theoretical background}
\begin{figure}[!thb]
\centering
\includegraphics[width=1.0\linewidth]{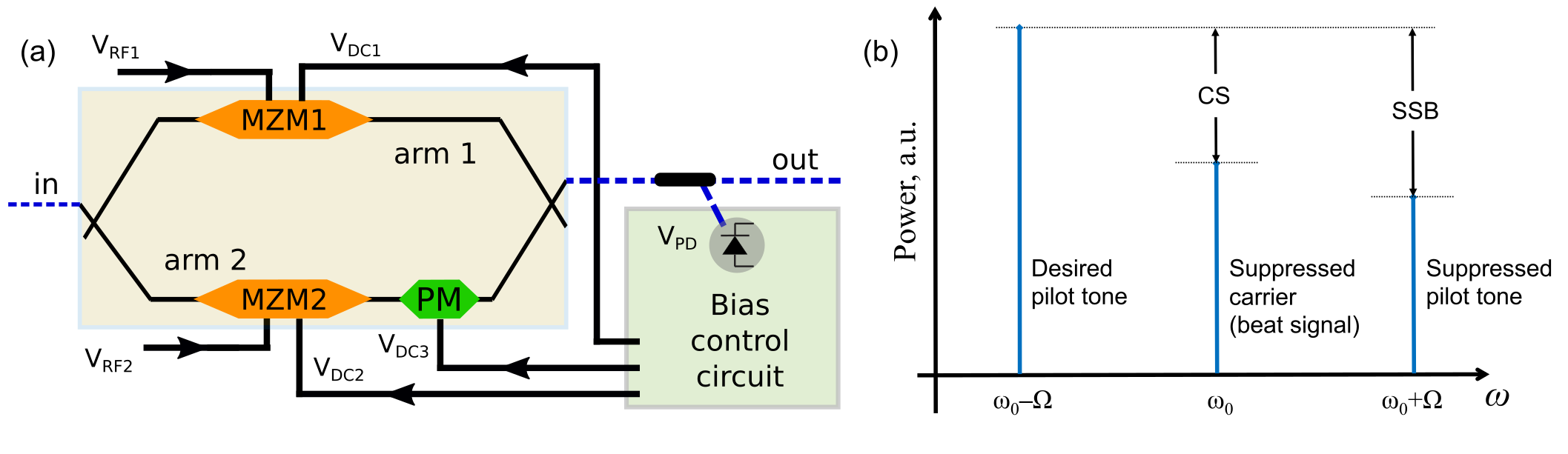}
\caption{Optical single sideband modulation with carrier suppression (OSSB-CS) from an IQ modulator. (a) The incoming light is split into two arms, and RF signals $V_{RF1}$ and $V_{RF2}$ with a phase difference of 90$^{\circ}$ drive Mach–Zehnder modulators (MZMs) operating on minima due to $V_{DC1}$ and $V_{DC2}$. A relative phase of 90$^{\circ}$, added by the phase modulator (PM) using $V_{DC3}$, then ensures OSSB-CS. The output is tapped to generate the photodiode signal $V_{PD}$ used for feedback control of the bias voltages. (b) Theoretical spectrum depicting both SSB and CS; also see equation~\eqref{eq:EoIQprac}. The amount of SSB and CS can be improved, though not always independently, by tuning the parameters such as dither amplitude and feedback photodiode gain of the bias circuit.}
\label{fig:setup}
\end{figure}
As shown in Fig.~\ref{fig:setup}(a), the IQ modulator is essentially a Mach-Zehnder interferometer consisting of two nested Mach-Zehnder modulators (MZMs) and a phase modulator (PM). These modulators are characterized by $V_{\pi}$, the voltage at which the optical phase changes by $\pi$. For a MZM, this means going from maximum optical transmission to the minimum, or vice versa. 

In the so-called optical single sideband modulation with carrier suppression (OSSB-CS) mode~\cite{Smith1997, Xue2014}, both MZM1 and MZM2 are biased at the minimum transmission point, e.g., $V_{DC1} = V_{DC2} = -V^{\rm MZM}_\pi$, while the PM voltage is e.g., $V_{DC3} = -V^{\rm PM}_{\pi/2}$ to make the optical signals at the output beam-splitter combine in quadrature. If the RF waveforms to the two MZMs are sinusoids in quadrature, such as $V_{RF1}(t) = A_1 \sin(\Omega t)$ and $V_{RF2}(t) = A_2 \cos(\Omega t)$, the electric field obtained at the output of (an ideal) IQ modulator is
\begin{equation}
E_o(t) \propto \left[ \sin \left(\mu_1\sin(\Omega t) \right) + i \sin \left(\mu_2\cos(\Omega t) \right) \right] E_i(t), 
\label{eq:EoIQapx1}
\end{equation}
given an electric field $E_i(t)$ at the input. Here,  $\mu_j = \pi A_j/2V^{\rm MZM}_{\pi}$ captures the effective modulation depth in arm $j$, for $j= 1 \text{ or } 2$. 

Using the (first two terms from) Jacobi-Anger expansion\footnote{$\sin (z\sin\theta) = 2\sum_{n=1}^{\infty} J_{2n-1}(z) \sin[(2n-1)\theta]$ and $\sin (z\cos\theta) = -2\sum_{n=1}^{\infty} (-1)^{n}J_{2n-1}(z) \cos[(2n-1)\theta]$.}, we can rewrite Eq.~\eqref{eq:EoIQapx1} as
\begin{eqnarray}
E_o(t) & \propto & \left[ J_1(\mu) \left(\cos(\Omega t) + i \sin(\Omega t) \right) - J_3(\mu) \left(\cos(3\Omega t) - i \sin(3\Omega t) \right) \right] E_i(t) \nonumber \\
& = & J_1(\mu) e^{i(\omega_0+\Omega) t} - J_3(\mu) e^{i(\omega_0-3\Omega) t} , 
\label{eq:EoIQ}
\end{eqnarray}
given an input field $E_i(t) \propto e^{i\omega_0 t}$ and assuming $\mu_1 \approx \mu_2 = \mu$. Here $J_k$ denotes a Bessel function of the first kind and order $k$. Note that a global phase of $\pi/2$ has been omitted in the above.

Eq.~\eqref{eq:EoIQ} illustrates a \textit{complete} suppression of the lower sideband at a frequency offset $-\Omega$ from the carrier. Moreover, since CVQKD transmitters operate the IQ modulator at low modulation depths for preserving linearity, one can consider $J_3(\mu) \approx 0$ because $\mu \ll 1$. The output field is then a single optical line at frequency $\omega_0+\Omega$, and exhibits perfect OSSB-CS. 

In practice, such an infinite suppression of the carrier and a sideband is however impossible because of imprecise DC biasing, finite manufacturing tolerance, RF mismatch, etc. In the above example, one can thereby anticipate the presence of a suppressed carrier at frequency $\omega_0$ and a suppressed sideband at $\omega_0-\Omega$ in the output field. Under the low modulation depth condition in Eq.~\eqref{eq:EoIQ} but with $\mu - \delta = \mu_1 \neq \mu_2 = \mu + \delta$, one obtains
\begin{align}
E_o(t) & \propto \left[ J_1(\mu_2)\cos(\Omega t) + i J_1(\mu_1)\sin(\Omega t) + \sin(\Delta_2) + i\sin(\Delta_1) \right] E_i(t) \nonumber \\
  &\approx \frac{1}{2}\left[ (\mu + \delta)\cos(\Omega t) + i (\mu - \delta)\sin(\Omega t) + \Delta_2 + i \Delta_1 \right] E_i(t) \nonumber \\
  &= \frac{\mu}{2} e^{i(\omega_0+\Omega) t} + \frac{\delta}{2} e^{i(\omega_0-\Omega) t}  + \Delta e^{i\omega_0 t}
\label{eq:EoIQprac}
\end{align}
on expanding the Bessel functions, with $\Delta_j$ denoting small deviations in the DC biases on the MZMs and $\Delta = \Delta_2 + i \Delta_1$ (by invoking $\sin(\Delta_j) \approx \Delta_j$). Figure~\ref{fig:setup}(b) illustrates a power spectrum that may represent the optical field in Eq.~\eqref{eq:EoIQprac}. The two sidebands, namely, the desired pilot tone and suppressed pilot tone, are located symmetrically around the suppressed beat signal. 
\subsection*{Leakage during state preparation} \label{bg:ldsp}
Coherent detection of the quantum data signal in CVQKD systems is performed by Bob, who now generally employs a locally generated `real' LO instead of using the `transmitted' LO from Alice. The sharing of phase reference across Alice and Bob is then done by so-called reference pulses (for pulsed systems) or pilot tones (for continuous-wave systems) \cite{Soh2015, Qu2016, Kleis2017, Brunner2017, Laudenbach2019, Chin2020}. We focus on the latter type, where a broadband signal is frequency multiplexed to the pilot tone, and then attenuated to yield the quantum data signal with a comparably bright pilot.

\begin{figure}[!t]
\centering
\includegraphics[width=1.0\linewidth]{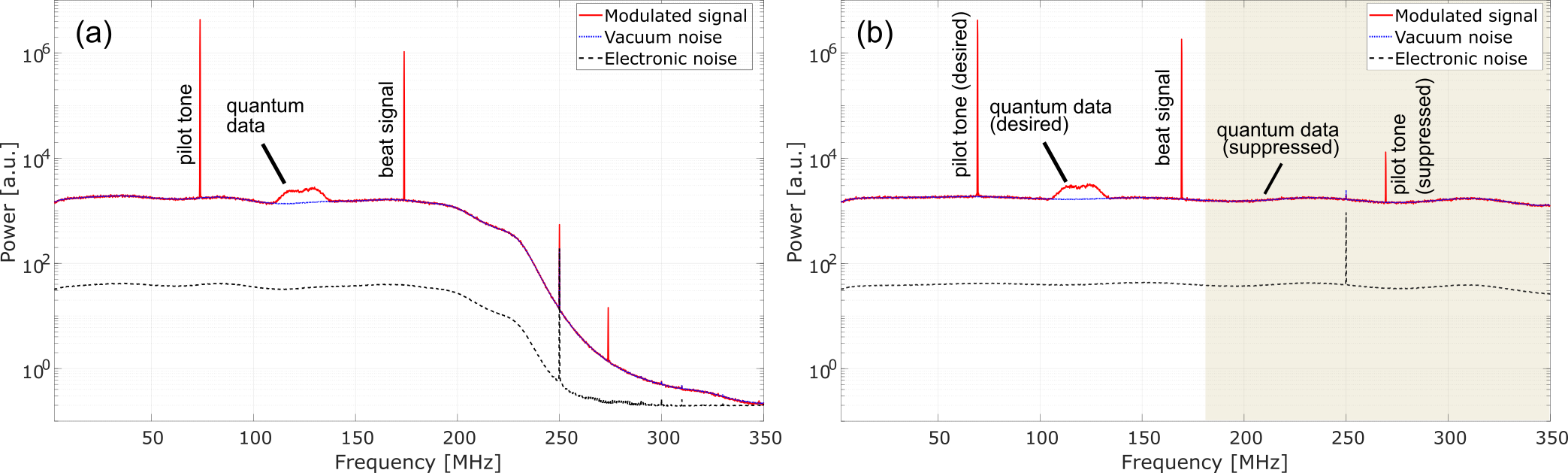}
\caption{Heterodyne spectra of frequency-multiplexed quantum data sideband and pilot tone (and other signals relevant to CVQKD measurements). (a) A low pass filter (LPF) with cutoff frequency of 200 MHz, used primarily for reducing out-of-band noise, also hides the suppressed sidebands quite well. (b) Once the 200 MHz LPF is removed, the suppressed pilot tone becomes much more apparent. The vulnerability arises if Alice and Bob do not take into account the leakage from the suppressed components (in the shaded region), as they are obviously present in the actual optical signal transmitted by Alice on the quantum channel, and thus fully accessible to Eve.}
\label{fig:quSignPilot}
\end{figure}

Figure~\ref{fig:quSignPilot} shows various spectra measured with a RF heterodyne detector in our CVQKD setup. 
We measure the detector electronic noise (dashed-black trace) when both the signal laser and LO are off, while with the signal laser off but LO on, we obtain vacuum noise (dotted-blue trace). With both light sources on and with the IQ modulator operating in OSSB-CS mode due to the DC bias control; see Fig.~\ref{fig:setup}(a), we observe the suppressed carrier / beat signal in the heterodyne spectra. On applying RF modulation, we obtain the modulated signal (solid-red trace) spectra that shows the two main sidebands on the left to the beat signal. 

In normal operation, we use a low pass filter (LPF) with cutoff around 200 MHz to limit Bob's detection bandwidth, as all relevant frequency components needed for carrier and phase recovery are present within this bandwidth~\cite{Chin2020}. The LPF is conspicuous by its absence in Fig.~\ref{fig:quSignPilot}(b): the suppressed pilot tone (SP) is fairly distinct in the spectra here. While the suppressed quantum band (SQB) signal may seem buried in the vacuum noise, it carries correlations with the signal in the desired quantum band (DQB) and can also be used for decoding the information by anyone having access to that sideband, albeit with some penalty.

Note that such a leakage does not affect the security of intradyne or phase-diverse CVQKD systems~\cite{Lance2005, Soh2015, Qu2016, Laudenbach2019} because Bob effectively measures both sidebands. However, even in such systems, the vulnerability can be present in case higher-order sidebands, produced, for example, due to a large modulation depth, lie outside the detection bandwidth of Bob. This may indeed happen in pulsed CVQKD transmitters that actually implement polar modulation~\cite{Jouguet2012}.
\subsection*{Attack model} %
\begin{figure}[!t]
            \centering
        \includegraphics[width=0.9\textwidth]{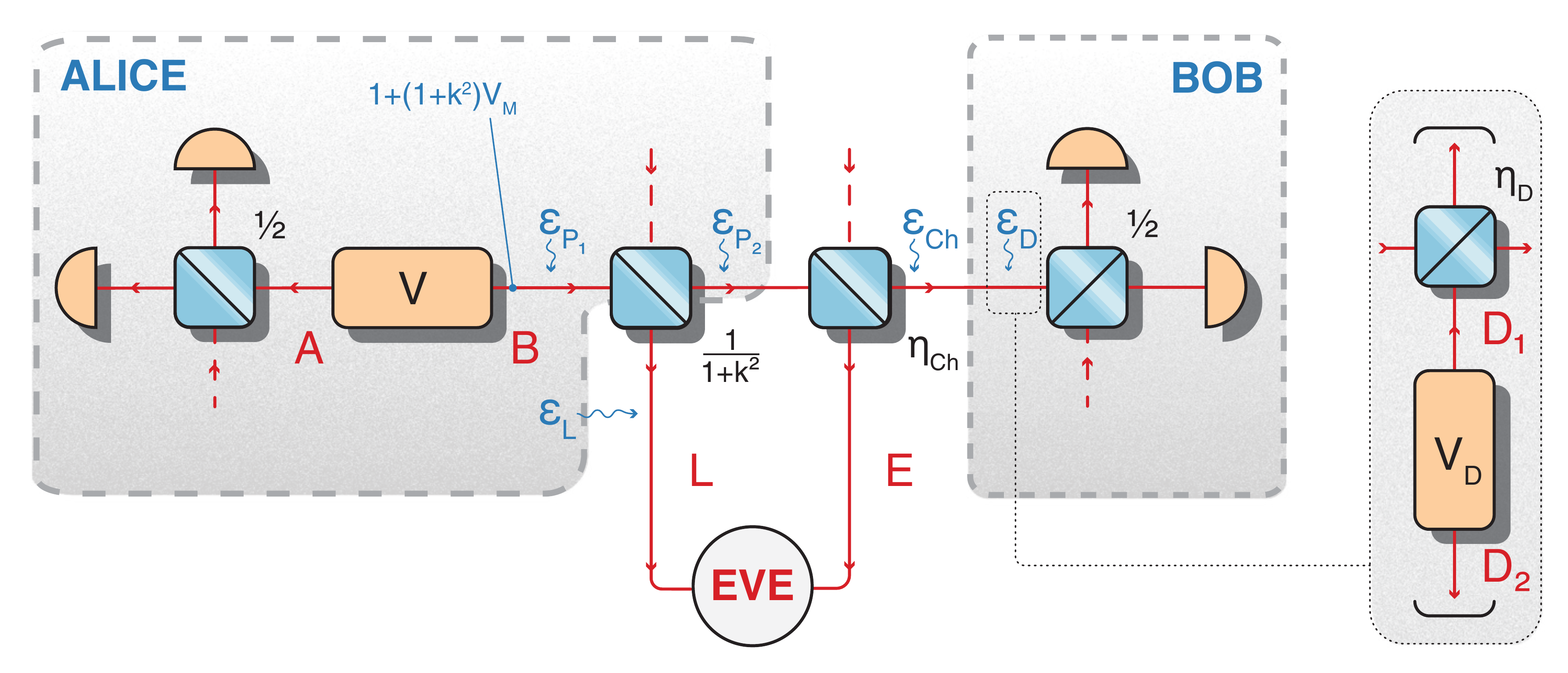}     \caption{Purification scheme used for security analysis. EPR source $V$ radiates states with quadrature variance $1+(1+k^2)V_M$ in each of modes A and B. We model the modulation leakage as a linear interaction of the signal (B) and leakage (L) modes on a beamsplitter with transmittance $1/(1+k^2)$. The signal is exposed to losses $\eta_{Ch}$ and excess noise with variance $\varepsilon_{Ch}$ in the untrusted channel. Eve obtains the information about the transmitted key from the output of the leakage mode (L) and from an auxiliary channel mode (E). Detection efficiency and trusted noise are modeled as a coupling with ratio $\eta_D$ of the signal mode entering Bob and mode $D_1$ of an entangled state with variance $V_D=1+\varepsilon_D/(1-\eta_D)$, where $\varepsilon_D$ is the variance of added detection noise. Possible infusion points of trusted preparation noise $\varepsilon_{P_1,P_2,L}$ are also shown. Trusted preparation noise is also modeled as a result of unbalanced coupling between the signal and an EPR source with appropriate variance.}
        \label{fig:scheme} 
\end{figure}
Modulation leakage in the coherent-state protocol can be considered as a Trojan-horse attack \cite{pereira2018}, where the vacuum state injected by Eve receives a fraction of the signal modulation. Such an attack is equivalent to the setup without the side channel leakage, but with the altered values of higher signal modulation $V_M'=(k^2+1)V_M$, and lower transmittance of the quantum channel $\eta'=\eta/(k^2+1)$, where $k^2=V_{M_L}/V_M$ is the ratio between variances of the leakage mode modulation and the signal (with $k\geq 0$), and the input of the leakage mode is assumed to be vacuum. The covariance matrix describing the effective two-mode state is \cite{pereira2018}:
    \begin{align}
        \gamma_{AB} = 
        \left(\begin{matrix}
        [1+(k^2+1)V_M]\mathbf{1} & \sqrt{\eta_{Ch} V_M[2 +(k^2+1)V_M]} \mathbf{P_3} \\ 
        \sqrt{\eta_{Ch} V_M[2 +(k^2+1)V_M]} \mathbf{P_3} & [1+\eta_{Ch} V_M + \varepsilon_{Ch}]\mathbf{1} \\
        \end{matrix}\right)\ ,
        \label{CM:AB}
    \end{align} 
where $\mathbf{1}$ is a $2\times 2$ identity matrix, $\mathbf{P_3}=\text{diag}[1,0,0,-1]$ is the Pauli matrix, and loss $\eta_{Ch}$ and excess noise with variance $\varepsilon_{Ch}$ characterize the untrusted channel. 

Figure~\ref{fig:scheme} shows the overall purification scheme for the attack model, with the entire state in 4 trusted modes $A,B,D_1$ and $D_2$, and Eve's modes $E$ and $L$.
The signal is subjected to trusted loss $\eta_D$ and trusted noise $\varepsilon_D$ stemming from the receiver, that are purified by an Einstein-Podolsky-Rosen (EPR) source radiating entangled states with variance $V_D=1+\varepsilon_D/(1-\eta_D)$ in modes $D_{1,2}$. Likewise, potential trusted preparation noise $\varepsilon_i$ ($i=P_1, P_2, L$) can be purified using a two-mode squeezed-vacuum source with variance $V_i=1+\varepsilon_i/(1-\eta_i)$ coupled to the signal on a strongly unbalanced beamsplitter $\eta_i\to 1$. Such noise can be applied either to the signal and leaked along with the modulation ($\varepsilon_{P_1}$), or to the signal only ($\varepsilon_{P_2}$). In practice, Eve's measurement of the leakage mode may be susceptible to limited detection efficiency and detection noise, limiting the channel advantage for Eve. We adopt a pessimistic approach and assume that during the experiment Eve could retrieve the leaked information with perfect efficiency, and purify and eliminate the noise at her side. Nevertheless, we discuss the influence of all possible types of noise~\cite{usenko2010feasibility}, including $\varepsilon_{L}$ associated to the leakage mode $L$, on the security in Sec. \ref{sec:results}.

Since Eve is assumed to hold a purification of the state shared between the trusted parties, one can use the equivalence between entropies of the state in Eve's mode and states in Alice and Bob modes \cite{braunstein2005quantum}. Hence a four-mode covariance matrix $\gamma_{ABD_{1,2}}$ is sufficient to evaluate the Holevo bound on the information accessible to Eve: 
\begin{equation}
       \chi_E^\text{DR}=S(ABD_{1,2})-S(BD_{1,2}|A)\ , \quad \chi_E^\text{RR}=S(ABD_{1,2})-S(AD_{1,2}|B)\ ,
\end{equation}
where $S(ABD_{1,2})$, $S(AD_{1,2}|B)$ and $S(BD_{1,2}|A)$ are the (conditional) Von Neumann entropies that are calculated based on the symplectic eigenvalues of respective covariance matrices $\gamma_{ABD_{1,2}}$, $\gamma_{AD_{1,2}|B}$ and $\gamma_{BD_{1,2}|A}$. The secret key fraction, in bits/symbol, for direct (DR,$\rightarrow$) and reverse (RR, $\leftarrow$) reconciliation is given by Ref.~\cite{holevo2001evaluating}:
\begin{equation}
    R^{\rightarrow(\leftarrow)}=\beta I_{AB}' - \chi_E^{\text{DR}(\text{RR})},
    \label{eq:keys}
\end{equation}
where $\beta \in [0,1]$ is the efficiency of the reconciliation algorithm, and $I_{AB}'$ is the mutual information\footnote{In the absence of trusted noise, $I_{AB}=I_{AB}^{x}+I_{AB}^{p}=\log_2\left[(1+\eta_{Ch} V_M)/(2+\varepsilon_{Ch})\right]$ for heterodyne detection.} modified according to the scheme depicted in Fig. \ref{fig:scheme}. For further details of the security estimation see Ref.~\cite{weedbrook2012gaussian}. 

\begin{figure}[t]
    \begin{tabular}{ll}
        \includegraphics[width=0.5\linewidth]{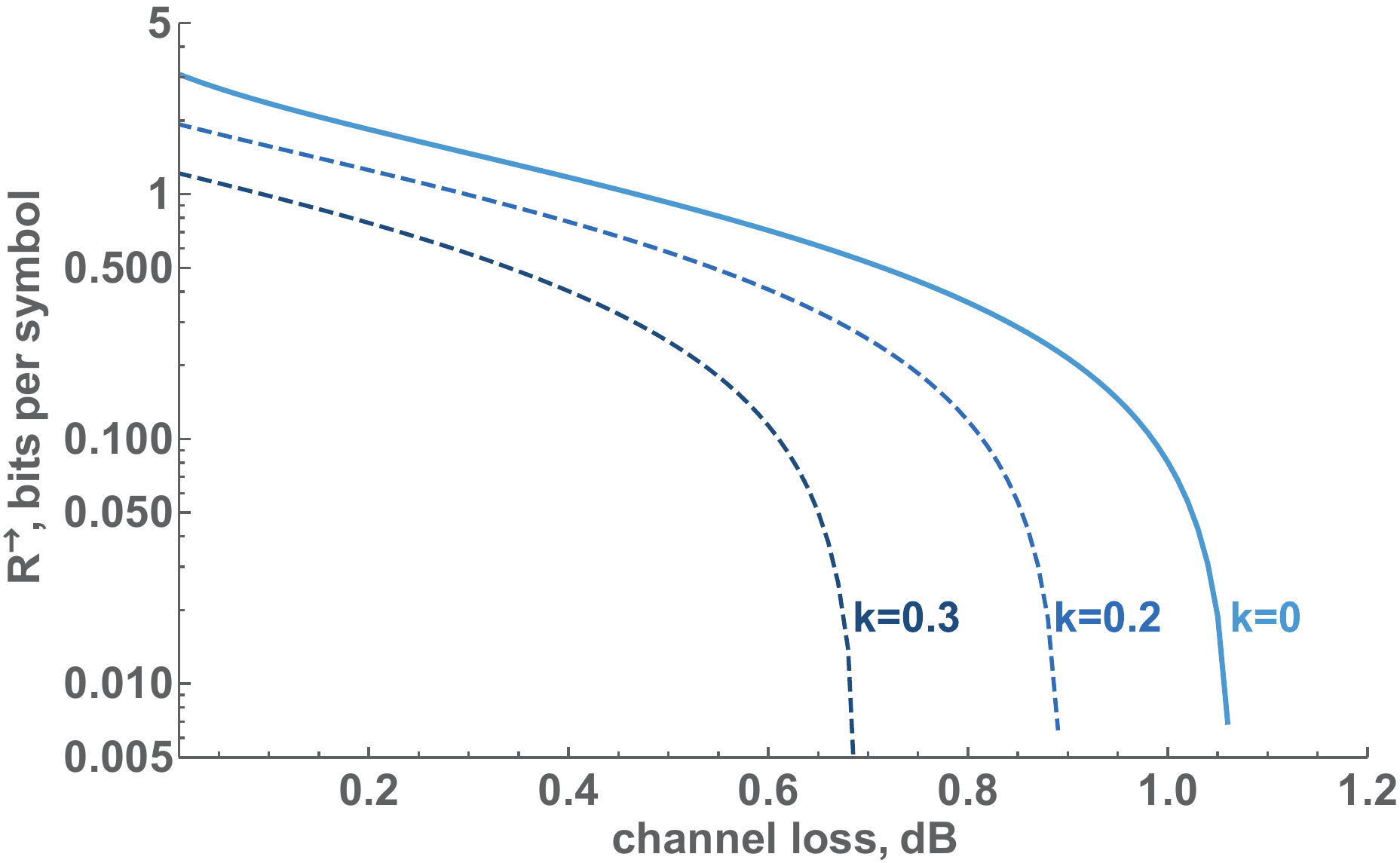}
        \includegraphics[width=0.5\linewidth]{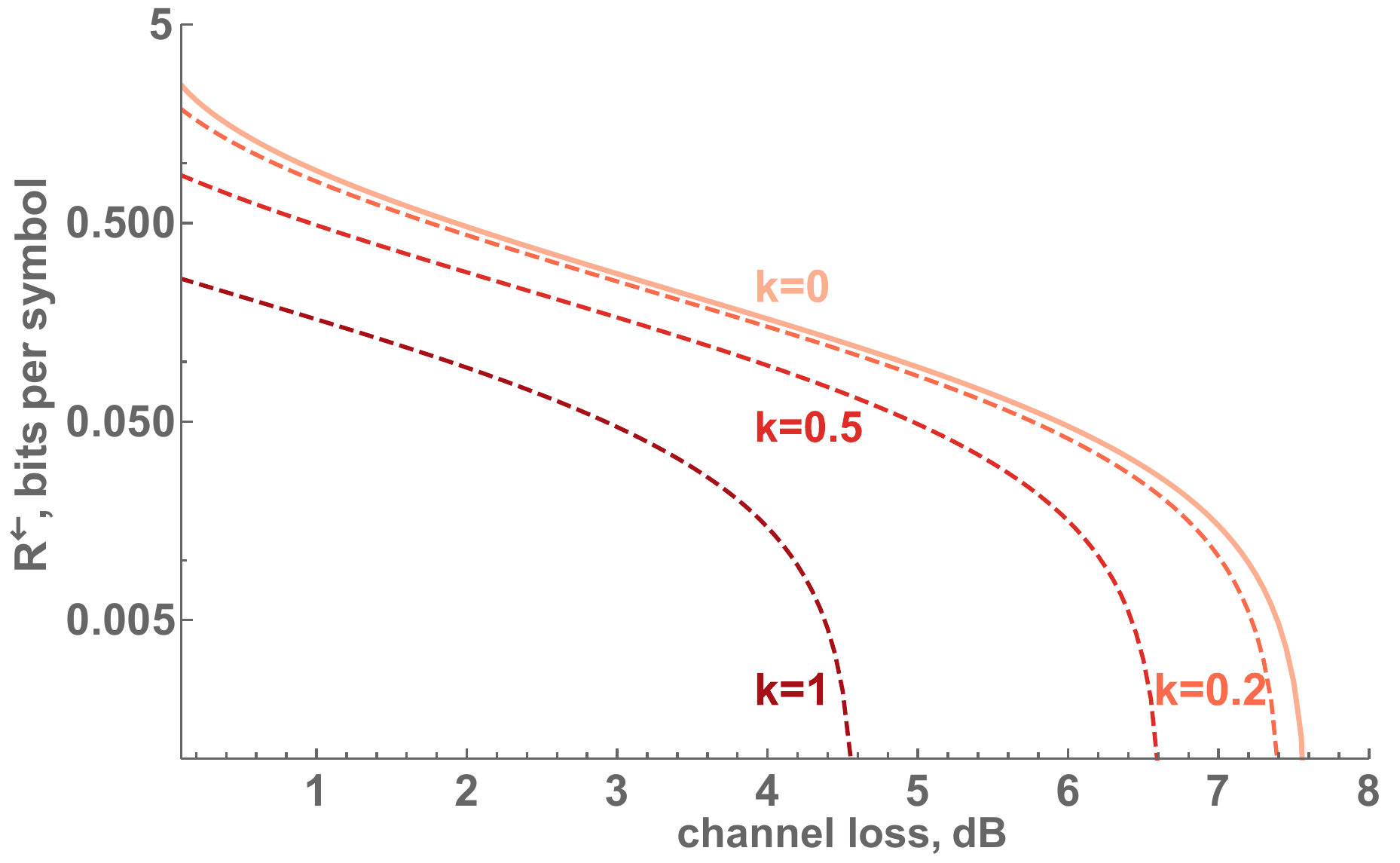}	
    \end{tabular}
    \caption{Lower bound on the secure key fraction (in bits per symbol) versus the channel loss (dB) for DR (left) and RR (right). Solid lines show the expected security in the absence of leakage, dashed lines show the influence of the modulation leakage with $k=0.2$ (-7.0 \text{dB}), 0.3 (-5.2 dB) for DR, and $k=0.2$ (-7.0 dB), 0.5 (-3.0 dB), 1 (0.0 dB) for RR. Reconciliation efficiency $\beta=0.96$, modulation variance $V_M$ is optimized, assumed untrusted excess noise at the channel output $\varepsilon_{Ch}= 0.02$ SNU. }
    \label{fig:key_on_eta}
\end{figure}

Figure~\ref{fig:key_on_eta} shows the resulting lower bounds of the secure key fraction of Eq.~(\ref{eq:keys}) as a function of the channel loss. 
As expected, CVQKD protocols adopting DR are very sensitive to modulation leakage and are not able to establish a secure key when the leakage ratio reaches $k=1$ \cite{Derkach2017}. Protocols with RR are less susceptible in comparison, however, with a diminished range of channel loss for secure operation~\cite{Derkach2017, pereira2018}. For typical sideband suppression ($>20\,$dB) in bulk modulators, the performance impact is rather small. However, we note that if not taken properly into account, the leakage results in a wrong evaluation of the lower bound on secure key fraction, and can lead to security breach at large channel loss.
\section{Experiment and Eve's attack strategy}
\begin{figure}[!t]
\centering
\includegraphics[width=1.0\linewidth]{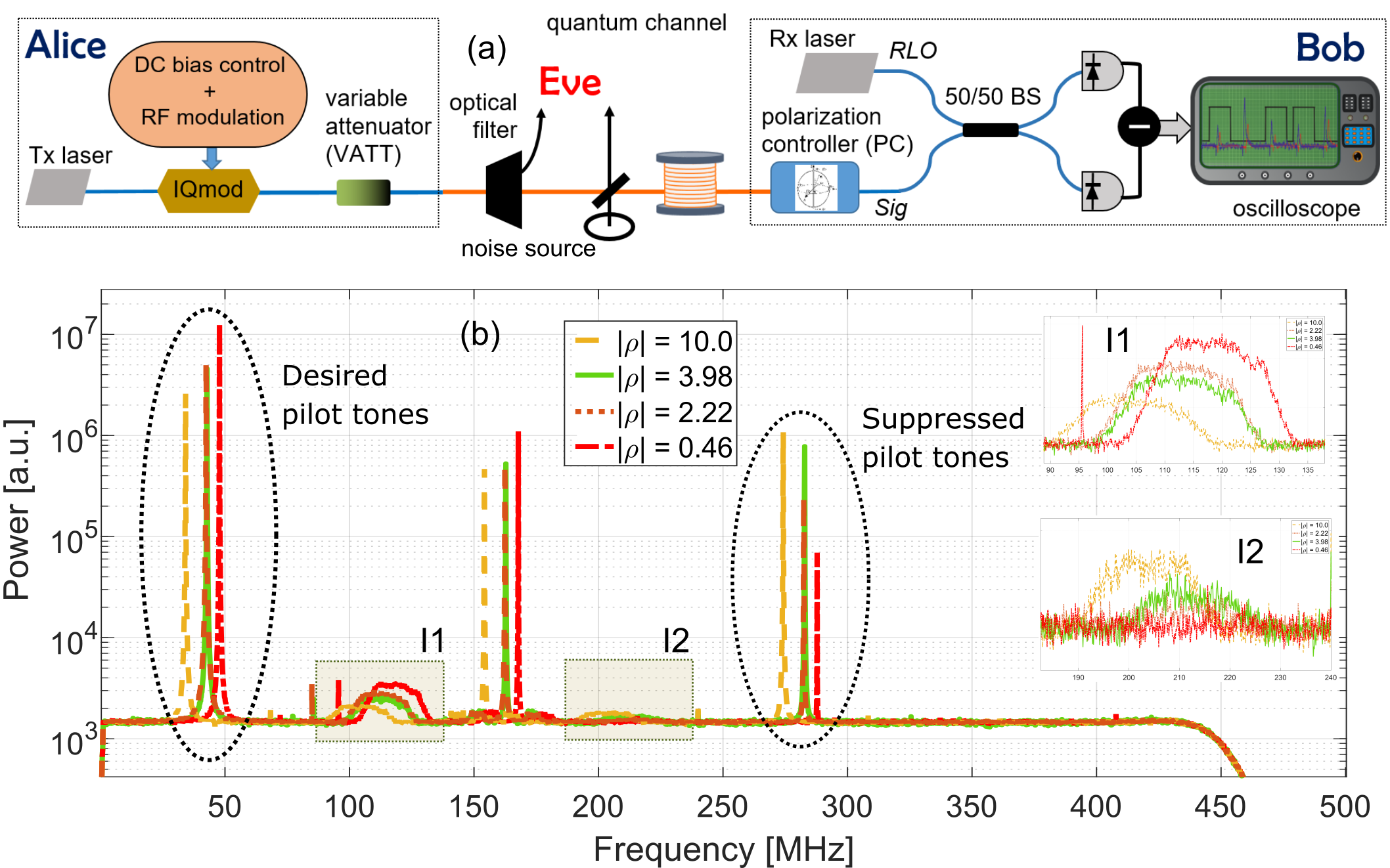}
\caption{Schematic of the experiment exposing the leakage vulnerability and measured heterodyne output spectra at varying degrees of sideband suppression. (a) Alice and Bob perform regular QKD measurements while Eve launches the attack illustrated in Fig.~\ref{fig:scheme} from the quantum channel. (b) Both the suppressed pilot and quantum data clearly become more apparent with the departure of the scaling $\rho$ from 0 dB. The insets I1 and I2 show the zoomed desired and suppressed quantum data bands, respectively.}
\label{fig:setupNbadSSB}
\end{figure}

Figure~\ref{fig:setupNbadSSB}(a) shows a simplified scheme of the prepare-and-measure CVQKD setup used for the experiment. 
We use a commercial off-the-shelf IQ modulator in the transmitter and a home-made broadband balanced detector in the receiver. The output of the transmitter (Tx) laser is modulated using RF waveforms prepared using an arbitrary waveform generator (not shown in the figure). The DC biases to the IQ modulator (IQmod) are controlled using a commercial automatic bias controller to obtain OSSB-CS. The modulated output is attenuated so that the quantum data band contains a few photons when it travels over the quantum channel to Bob. 

Using a manual polarization controller, we optimize the received signal's polarization for RF heterodyne detection with a real LO, generated by the receiver (Rx) laser. An oscilloscope samples and acquires the balanced detector output. The acquired data is used to reconstruct Alice's data using various digital signal processing methods, in particular, a machine learning framework based on Bayesian inference, for highly accurate phase estimation and compensation~\cite{Chin2020}. 

Since information is (also) encoded in the suppressed sidebands, Eve can use an optical filter, such as an optical add-drop multiplexer (OADM) as exhibited in Fig.~\ref{fig:setup}(a)), to divert a part of the spectrum to herself, while transmitting the rest to Bob. If Eve intercepts the SQB (shaded region I2 in Fig.~\ref{fig:setup}(b)) and if Alice and Bob do not take this into account, then the security of the final key can be compromised. 

In the experiment, we connected Alice and Bob without any channel, i.e., in a back-to-back (B2B) configuration. From the acquired data, we processed the desired and suppressed quantum data bands (see Fig.~\ref{fig:quSignPilot}(b)) \textit{independently} of each other. The sharing of phase reference, i.e., phase corrections to the quantum data, can be performed using either the desired or the suppressed pilot tone. resulting in the following possible measurement strategies for Eve: 
\begin{itemize}
    \item SQB-SP: Suppressed quantum band processed using suppressed pilot tone, and
    \item SQB-DP: Suppressed quantum band processed using desired pilot tone,
\end{itemize}
while Bob's measurement involves processing of the desired quantum band using the desired pilot tone (DQB-DP). Eve's second strategy can be justified on the basis that DP is a classical signal, and therefore, Eve can manipulate -- intercept, utilize, prepare afresh and re-send -- it without any eventual penalty. 

To highlight the vulnerability, we performed regular CVQKD measurements while varying the amount of sideband suppression by scaling the two RF output voltage levels that drive the IQ modulator with respect to each other; see  Fig.~\ref{fig:setup}(a). Eve is assumed to access the relevant parts of the spectrum using a perfect OADM, as illustrated in Fig.~\ref{fig:setupNbadSSB}(a). The RF scaling factor, given by $\rho = 10\log\frac{\text{max}(V_{\text{RF}1}(t))}{\text{max}(V_{\text{RF}2}(t))}$, is expressed in dB. We processed the acquired data from the heterodyne detector on a frame-to-frame basis, with each frame consisting of $10^7$ samples from the oscilloscope. We acquired 20 frames per value of $\rho$ for statistics. 

Figure~\ref{fig:setupNbadSSB}(b) shows the spectra from modulated frames acquired at four different values of $\rho$. The averaged frequency response obtained from the power spectra of the measured vacuum noise (see the dotted-blue trace in Fig.~\ref{fig:quSignPilot}(b) for example), was inverted to create a `whitening' filter. Applying this filter to the acquired data frames results in the flat response from near DC to $>400$ MHz. To process the signal of interest, we performed carrier recovery with the help of a machine learning framework that employs an Unscented Kalman Filter (UKF)~\cite{Chin2020}. 
\section{Results and Discussion}\label{sec:results}
\begin{figure}[p!]
    \centering
    \includegraphics[width=.85\textwidth]{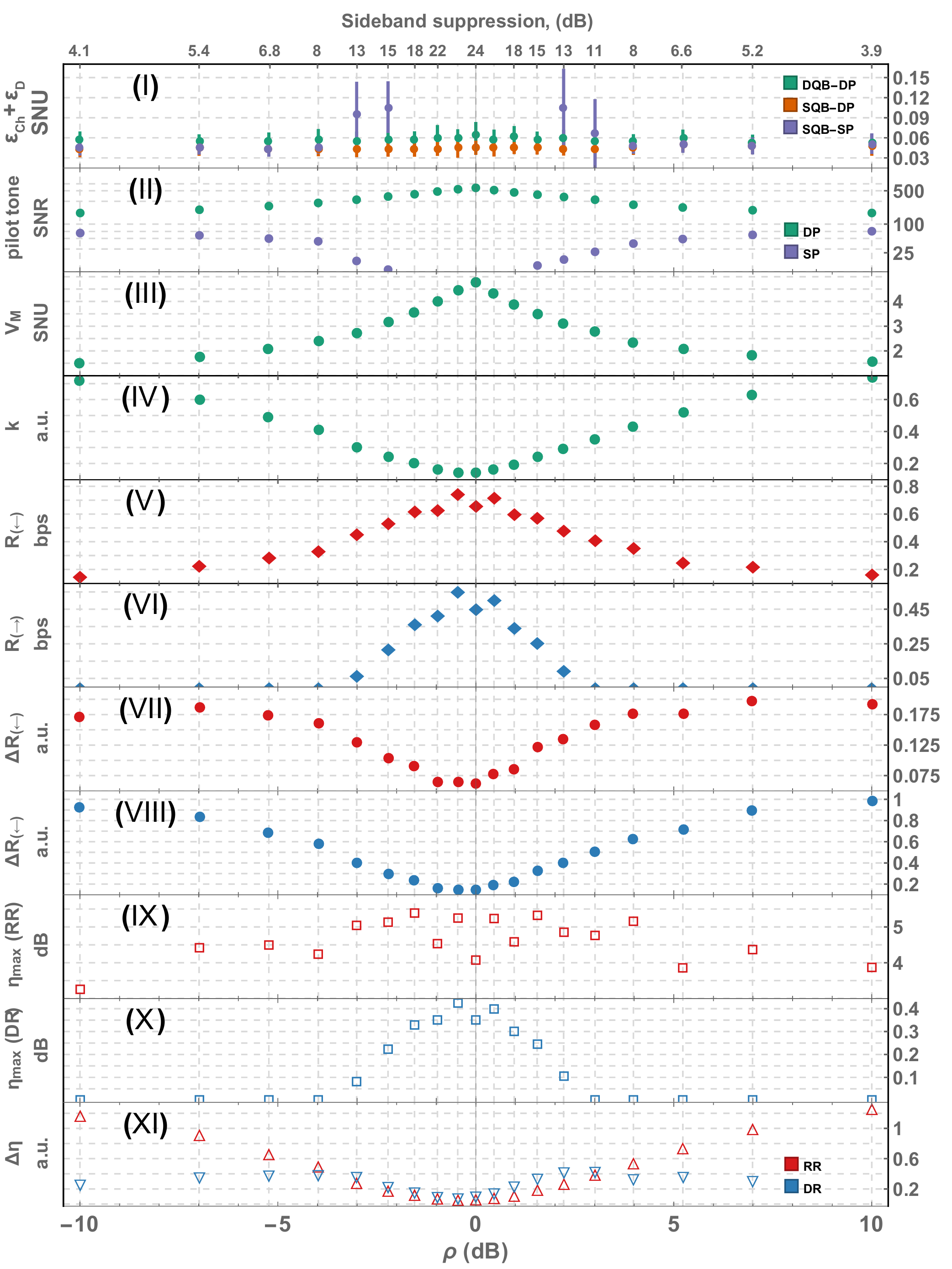}
    \caption{Values of the estimated parameters (i-iv) and security analysis results (v-xi) as a function of the RF scaling $\rho$ (bottom X axis). The corresponding sideband suppression values, estimated as the ratio of desired to suppressed pilot tone powers, are presented on the top X axis. From top to bottom: (i) the estimated total excess noise in shot noise units (SNU) from the processing of the SQB and DQB; (ii) signal-to-noise ratio (SNR) of desired and suppressed pilot tones; (iii) modulation variance $V_M$ obtained from DQB; and (iv) leakage parameter $k$. The security analysis provides a lower bound on the key fraction in bits per symbol (bps) with RR (v) and DR (vi); difference of secure key fraction $\Delta R$ without leakage and with side channel leakage $k\neq 0$ for RR (vii) and DR (viii) techniques; maximal tolerable additional channel loss $\eta_{max}$ in dB for RR and DR (ix) and (x) respectively; (xi) difference between maximal tolerable additional channel loss provided trusted parties are ignorant about modulation leakage ($k=0$) and if the leakage is taken into account $\Delta\eta$ for RR (red) and DR (blue) techniques.}
    \label{fig:results}
    \end{figure}

Figure~\ref{fig:results}(i) shows the total excess noise $\varepsilon_{Ch}+\varepsilon_D$ versus the RF scaling factor $\rho$ for the 3 different data processing strategies mentioned before. 
Here, negative [positive] $\rho$ values correspond to varying the peak-to-peak voltage of the applied waveform on arm 1 [arm 2] while keeping the peak-to-peak voltage on arm 2 [arm 1] constant; see Fig. \ref{fig:setup}(a). For poor sideband suppression ($|\rho|>4\,$dB), the average value of $\varepsilon_{Ch}<0.06$ SNU, regardless of whichever pilot tone is used by the UKF for the purpose of phase recovery. For $|\rho|<4\,$dB, the suppressed pilot tone power is not large enough to provide enough SNR for a proper carrier recovery: consequently, the excess noise increases rapidly. Note that the green curve is above the orange curve because the modulation variance is higher for DQB compared to the SQB. 

The pilot tone SNR values available to UKF are plotted in Fig.~\ref{fig:results}(ii); for $\text{SNR}<10$, we could not obtain a sufficient number of processed frames with decent correlations for a reasonable estimation of the excess noise. Nonetheless, we note that in our experiment, we use much less powerful pilot tones compared to most other CVQKD setups \cite{Kleis2017, Brunner2017, Laudenbach2019}. So in general, there is a good likelihood that even the suppressed pilot tones provide a reasonable SNR for successful reconstruction. Finally, the slight asymmetry across the $|\rho|=0\,$dB vertical line stems from experimental imperfections, e.g., the mismatch in the electrical-to-optical response at the two MZMs in the IQ modulator.

Evaluating Bob's (DQB-DP) and Eve's (SQB-DP) data, we obtained a map between $\rho$ and the modulation variance $V_M$($\rho$) and the leakage parameter $k$($\rho$). These two parameters are depicted in Fig.~\ref{fig:results}(iii) and (iv), respectively, and along with the upper bound estimates of $\varepsilon_{Ch}(\rho)$ and $\eta_{Ch}(\rho)$, are used to construct the covariance matrix $\gamma_{ABD_{1,2}}(\rho)$ and assess the lower bound on the key fraction in Eq.~\eqref{eq:keys} subsequently. The secure key fractions in B2B configuration for RR (red) and DR (blue) obtained with a reconciliation efficiency $\beta=96\%$ and finite-size effects for a block size of $10^7$ \cite{Leverrier2010} are shown in Fig.~\ref{fig:results} (v) and (vi), respectively. Evidently, the best sideband suppression yields the highest values of signal modulation variance $V_M$ and lowest amount of leakage $k$, which consequently translates into higher levels of achievable secret key. Note that $k$ never reaches zero due to the experimental inability to completely eliminate the suppressed components. The protocol based on DR is as expected very sensitive to the modulation leakage \cite{Derkach2017} and cannot deliver a secret key with poor sideband suppression, i.e. $|\rho|>3\, \text{dB}$. The RR technique, on the other hand, can tolerate more leakage and can securely operate regardless of the sideband suppression, although with significantly reduced key fraction.

The sensitivity of the CVQKD system to leakage is also highlighted in Fig.~\ref{fig:results}(vii) and (viii), where Eve's information advantage, i.e.,\ the difference $\Delta R=R(k= 0)-R(k \neq 0)$ between the secret key fractions, is shown for RR and DR, respectively. One way to interpret this would be, if Alice and Bob are ignorant of the side channel leakage they would be generating a key at rate $\Delta R$ higher than what is actually secure. 

As the main sources of loss and noise are taken into account, we assess the maximal tolerable \textit{additional} loss $\eta_{max}^{(\rightarrow,\leftarrow)}$ of the untrusted channel, i.e., the total loss is given by $\eta_{Ch}\eta_{max}^{(\rightarrow,\leftarrow)}$. The results are shown in Fig.~\ref{fig:results}(ix) for RR and (x) for DR techniques. For RR  $\eta_{max}^{(\leftarrow)}$ is largely influenced by the noise encountered by Bob and Eve; see Fig.~\ref{fig:results}(i), while for DR, this additional loss is again determined by the amount of information leakage $k$. By ignoring the leakage, trusted parties might assume secure operation with up to $\eta_{max}^{(\rightarrow,\leftarrow)}(k=0)$ of additional channel loss, however the actual maximal loss, as shown in Fig.~\ref{fig:results}(ix) and (x), is lower. This is highlighted in Fig.~\ref{fig:results}(xi). Here $\Delta\eta=\eta_{max}^{(\rightarrow,\leftarrow)}(k=0)-\eta_{max}^{(\rightarrow,\leftarrow)}(k\neq0)$ shows the regime where ignorant Alice and Bob would assume secure key distribution, while in fact the RR (red) or DR (blue) based CVQKD protocol is not secure anymore.

A possible approach to improve the secure key length under modulation leakage involves injection of trusted noise to the reference side~\cite{usenko2010feasibility, Usenko2016}. In our experiment, preparation noise was not characterized and was thus attributed to the untrusted channel, whereas detection noise was accurately identified. Such trusted noise on either side can be controlled and optimized in order to improve or even recover the security of CVQKD protocol in a noisy untrusted channel. However, it is important to recognize and characterize the trusted noise and identify where it is referred to, as it can affect only the signal or both the signal and leaked modulation (see Fig. \ref{fig:scheme}) and carry different repercussions for security. 

We identify the conditions under which trusted noise can be used to improve the key rate and summarize the results in Table \ref{table:trusted-noise}.
\begin{table}
\centering
\begin{tabular}{ r | c | c } 
Trusted noise: & DR & RR \\
  \hline\hline 
Signal \& leakage $\varepsilon_{P_1}$ & \cmark &  \xmark \\
Signal only $\varepsilon_{P_2}$  & \cmark &  \xmark\\
Leakage $\varepsilon_{L}$  & \cmark &  \cmark\\
Detection $\varepsilon_{D}$  & \xmark &  \cmark \\ 
[1ex]
\end{tabular}
\caption{Viability of positive influence of trusted noise on the security of the coherent-state CVQKD protocol with modulation leakage.}
\label{table:trusted-noise}
\end{table}
Firstly, any trusted noise $\varepsilon_L$ in the leakage mode $L$ will translate into improved secure key fraction. As the DR technique is more sensitive to leakage (compared to RR) it will also benefit more from the respective noise, especially for stronger leakage. Secondly, the effect of trusted detection noise $\varepsilon_{D}$ is not altered by the leakage mode, and can be defensively used during RR only~\cite{Usenko2016}. Lastly, overall influence of preparation noise remains similar to conventional CVQKD operation without the leakage. For DR, any preparation noise ($\varepsilon_{P_1}$ or $\varepsilon_{P_2}$) can be used to the advantage of Alice and Bob. Although, if the noise is leaked along with the signal ($\varepsilon_{P_1}$) it can also directly hinder the effect of the leakage. For RR, trusted preparation noise can actually be harmful, even though $\varepsilon_{P_1}$ seemingly curtails the usefulness of that leakage to Eve. 
\section{Conclusion}
Optical single sideband (OSSB) encoding is a well-known technique in classical optical communication. It also has the potential to revolutionize broadband continuous-variable quantum key distribution (CVQKD) protocols by offering very low excess noise performance. OSSB modulation requires the suppression of a sideband, which can readily be implemented using an optical in-phase and quadrature (IQ) modulator. However, the amount of suppression is limited in practice, and as we have shown here, this can lead to the modulation leakage vulnerability in CVQKD systems. We have also presented a theoretical framework that analyses the insecurity resulting from this vulnerability: While the reverse reconciliation strategy suffers a reduction of the secret key length that becomes significant at higher leakages, the direct reconciliation strategy cannot produce a secret key even at moderate leakages. As a countermeasure, we have shown that depending on the type of reconciliation adopted, the trusted parties performing the CVQKD protocol could use preparation or detection noise, which is not in control of the adversary, to reduce the severity of the leakage. Finally, we note that such a leakage is much more likely in (future) photonic integrated circuit based modulators compared to bulk modulators. 
With IQ modulators poised to become the workhorses of CVQKD systems, we therefore believe this study is timely and can help protecting future CVQKD implementations against this vulnerability. 
\section*{Acknowledgements}
We acknowledge financial support from European Union's Horizon 2020 research and innovation programmes CiViQ (grant agreement no.\ 820466), OPENQKD (grant agreement no.\ 857156), and CSA Twinning NONGAUSS (grant agreement no.\ 951737). NJ, HMC, ULA, and TG acknowledge support from Innovation Fund Denmark (CryptQ project, grant agreement no.\ 0175-00018A) and the Danish National Research Foundation, Center for Macroscopic Quantum States (bigQ, DNRF142). ID and VCU acknowledge support from the project 19-23739S of the Czech Science Foundation. RF acknowledges support of Horizon 2020 Framework Programme (731473, project 8C20002 ShoQC). 
\bibliography{lib}

\begin{thebibliography}{10}
\newcommand{\enquote}[1]{``#1''}

\bibitem{Griffin02}
R.~Griffin and A.~Carter, \enquote{Optical differential quadrature phase-shift
  key (odqpsk) for high capacity optical transmission,} in \emph{Optical Fiber
  Communications Conference,}  (Optical Society of America, 2002), p. WX6.

\bibitem{Kikuchi2010}
K.~Kikuchi, \enquote{{Coherent Optical Communications: Historical Perspectives
  and Future Directions},} in \emph{Electrical Engineering,}  M.~Nakazawa,
  K.~Kikuchi, and T.~Miyazaki, eds. (Springer, Berlin, Heidelberg, 2010).

\bibitem{Ralph99}
T.~C. Ralph, \enquote{Continuous variable quantum cryptography,}
  {\protect\JournalTitle{Phys. Rev. A}} \textbf{61}, 010303 (1999).

\bibitem{Grosshans02}
F.~Grosshans and P.~Grangier, \enquote{Continuous variable quantum cryptography
  using coherent states,} {\protect\JournalTitle{Phys. Rev. Lett.}}
  \textbf{88}, 057902 (2002).

\bibitem{bennett1984}
C.~H. Bennett and G.~Brassard, \enquote{{Quantum Cryptography: Public Key
  Distribution and Coin Tossing},} in \emph{Proceedings of IEEE International
  Conference on Computers, Systems, and Signal Processing,}  (Bangalore, India,
  1984), pp. 175--179.

\bibitem{Scarani2009}
V.~Scarani, H.~Bechmann-Pasquinucci, N.~J. Cerf, M.~Du{\v{s}}ek,
  N.~L{\"{u}}tkenhaus, and M.~Peev, \enquote{{The security of practical quantum
  key distribution},} {\protect\JournalTitle{Reviews of Modern Physics}}
  \textbf{81}, 1301--1350 (2009).

\bibitem{Diamanti2015}
E.~Diamanti and A.~Leverrier, \enquote{{Distributing secret keys with quantum
  continuous variables: Principle, security and implementations},}
  {\protect\JournalTitle{Entropy}} \textbf{17}, 6072--6092 (2015).

\bibitem{Pirandola2020}
S.~Pirandola, U.~L. Andersen, L.~Banchi, M.~Berta, D.~Bunandar, R.~Colbeck,
  D.~Englund, T.~Gehring, C.~Lupo, C.~Ottaviani, J.~L. Pereira, M.~Razavi,
  J.~S. Shaari, M.~Tomamichel, V.~C. Usenko, G.~Vallone, P.~Villoresi, and
  P.~Wallden, \enquote{Advances in quantum cryptography,}
  {\protect\JournalTitle{Adv. Opt. Photon.}} \textbf{12}, 1012--1236 (2020).

\bibitem{BachorRalph}
H.~Bachor and T.~Ralph, \emph{A Guide to Experiments in Quantum Optics} (Wiley,
  2019).

\bibitem{Lance2005}
A.~M. Lance, T.~Symul, V.~Sharma, C.~Weedbrook, T.~C. Ralph, and P.~K. Lam,
  \enquote{{No-Switching Quantum Key Distribution Using Broadband Modulated
  Coherent Light},} {\protect\JournalTitle{Physical Review Letters}}
  \textbf{95}, 180503 (2005).

\bibitem{Jain2016}
N.~Jain, B.~Stiller, I.~Khan, D.~Elser, C.~Marquardt, and G.~Leuchs,
  \enquote{{Attacks on practical quantum key distribution systems (and how to
  prevent them)},} {\protect\JournalTitle{Contemporary Physics}} \textbf{57},
  366--387 (2016).

\bibitem{Jouguet2012}
P.~Jouguet, S.~Kunz-Jacques, E.~Diamanti, and A.~Leverrier, \enquote{{Analysis
  of imperfections in practical continuous-variable quantum key distribution},}
  {\protect\JournalTitle{Physical Review A}} \textbf{86}, 1--9 (2012).

\bibitem{Ma2013}
X.~C. Ma, S.~H. Sun, M.~S. Jiang, and L.~M. Liang, \enquote{{Local oscillator
  fluctuation opens a loophole for Eve in practical continuous-variable
  quantum-key-distribution systems},} {\protect\JournalTitle{Physical Review
  A}} \textbf{88}, 1--7 (2013).

\bibitem{Jouguet2013}
P.~Jouguet, S.~Kunz-Jacques, and E.~Diamanti, \enquote{{Preventing calibration
  attacks on the local oscillator in continuous-variable quantum key
  distribution},} {\protect\JournalTitle{Physical Review A}} \textbf{87}, 1--6
  (2013).

\bibitem{Stiller2015}
B.~{Stiller}, I.~{Khan}, N.~{Jain}, P.~{Jouguet}, S.~{Kunz-Jacques},
  E.~{Diamanti}, C.~{Marquardt}, and G.~{Leuchs}, \enquote{{Quantum hacking of
  continuous-variable quantum key distribution systems: Realtime Trojan-horse
  attacks},} in \emph{Conference on Lasers and Electro-Optics (CLEO),}  (2015),
  pp. 1--2.

\bibitem{Qin2016}
H.~Qin, R.~Kumar, and R.~All{\'{e}}aume, \enquote{{Quantum hacking: Saturation
  attack on practical continuous-variable quantum key distribution},}
  {\protect\JournalTitle{Physical Review A}} \textbf{94}, 012325 (2016).

\bibitem{Zhao2019}
Y.~Zhao, Y.~Zhang, Y.~Huang, B.~Xu, S.~Yu, and H.~Guo, \enquote{{Polarization
  attack on continuous-variable quantum key distribution},}
  {\protect\JournalTitle{Journal of Physics B: Atomic, Molecular and Optical
  Physics}} \textbf{52} (2019).

\bibitem{Derkach2017}
I.~Derkach, V.~C. Usenko, and R.~Filip, \enquote{{Continuous-variable quantum
  key distribution with a leakage from state preparation},}
  {\protect\JournalTitle{Physical Review A}} \textbf{96}, 062309 (2017).

\bibitem{Smith1997}
G.~Smith, D.~Novak, and Z.~Ahmed, \enquote{{Technique for optical SSB
  generation to overcome dispersion penalties in fibre-radio systems},}
  {\protect\JournalTitle{Electronics Letters}} \textbf{33}, 74 (1997).

\bibitem{Chin2020}
H.-M. Chin, N.~Jain, D.~Zibar, U.~L. Andersen, and T.~Gehring,
  \enquote{{Machine learning aided carrier recovery in continuous-variable
  quantum key distribution},} {\protect\JournalTitle{npj Quantum Information}}
  \textbf{7}, 20 (2021).

\bibitem{Qu2016}
Z.~Qu, I.~B. Djordjevic, and M.~A. Neifeld, \enquote{{RF-subcarrier-assisted
  four-state continuous-variable QKD based on coherent detection},}
  {\protect\JournalTitle{Optics Letters}} \textbf{41}, 5507 (2016).

\bibitem{Kleis2017}
S.~Kleis, M.~Rueckmann, and C.~G. Schaeffer, \enquote{{Continuous variable
  quantum key distribution with a real local oscillator using simultaneous
  pilot signals},} {\protect\JournalTitle{Optics Letters}} \textbf{42},
  1588--1591 (2017).

\bibitem{Brunner2017}
H.~H. Brunner, L.~C. Comandar, F.~Karinou, S.~Bettelli, D.~Hillerkuss, F.~Fung,
  D.~Wang, S.~Mikroulis, M.~Kuschnerov, A.~Poppe, C.~Xie, and M.~Peev,
  \enquote{{Low-noise, low-complexity CV-QKD architecture},} in \emph{QCrypt
  2017,}  (Cambridge, 2017), pp. 2--4.

\bibitem{Laudenbach2019}
F.~Laudenbach, B.~Schrenk, C.~Pacher, M.~Hentschel, C.-H.~F. Fung, F.~Karinou,
  A.~Poppe, M.~Peev, and H.~H{\"{u}}bel, \enquote{Pilot-assisted intradyne
  reception for high-speed continuous-variable quantum key distribution with
  true local oscillator,} {\protect\JournalTitle{{Quantum}}} \textbf{3}, 193
  (2019).

\bibitem{Jain2014}
N.~Jain, B.~Stiller, I.~Khan, V.~Makarov, C.~Marquardt, and G.~Leuchs,
  \enquote{{Risk analysis of Trojan - horse attacks on practical quantum key
  distribution systems},} {\protect\JournalTitle{IEEE Journal on Selected Topic
  in Quatum Electronics}} \textbf{21}, 1077--260X (2014).

\bibitem{Xue2014}
M.~Xue, S.~Pan, and Y.~Zhao, \enquote{{Optical single-sideband modulation based
  on a dual-drive MZM and a 120° hybrid coupler},}
  {\protect\JournalTitle{Journal of Lightwave Technology}} \textbf{32},
  3317--3323 (2014).

\bibitem{Soh2015}
D.~B.~S. Soh, C.~Brif, P.~J. Coles, N.~Luetkenhaus, R.~M. Camacho, J.~Urayama,
  and M.~Sarovar, \enquote{{Self-referenced continuous-variable quantum key
  distribution protocol},} {\protect\JournalTitle{Physical Review X}}
  \textbf{5}, 1--15 (2015).

\bibitem{pereira2018}
J.~Pereira and S.~Pirandola, \enquote{{Hacking Alice's box in
  continuous-variable quantum key distribution},}
  {\protect\JournalTitle{Physical Review A}} \textbf{98}, 062319 (2018).

\bibitem{usenko2010feasibility}
V.~C. Usenko and R.~Filip, \enquote{Feasibility of continuous-variable quantum
  key distribution with noisy coherent states,} {\protect\JournalTitle{Physical
  Review A}} \textbf{81}, 022318 (2010).

\bibitem{braunstein2005quantum}
S.~L. Braunstein and P.~Van~Loock, \enquote{Quantum information with continuous
  variables,} {\protect\JournalTitle{Reviews of Modern Physics}} \textbf{77},
  513 (2005).

\bibitem{holevo2001evaluating}
A.~S. Holevo and R.~F. Werner, \enquote{Evaluating capacities of bosonic
  gaussian channels,} {\protect\JournalTitle{Physical Review A}} \textbf{63},
  032312 (2001).

\bibitem{weedbrook2012gaussian}
C.~Weedbrook, S.~Pirandola, R.~Garc{\'\i}a-Patr{\'o}n, N.~J. Cerf, T.~C. Ralph,
  J.~H. Shapiro, and S.~Lloyd, \enquote{Gaussian quantum information,}
  {\protect\JournalTitle{Reviews of Modern Physics}} \textbf{84}, 621 (2012).

\bibitem{Leverrier2010}
A.~Leverrier, F.~Grosshans, and P.~Grangier, \enquote{Finite-size analysis of a
  continuous-variable quantum key distribution,}
  {\protect\JournalTitle{Physical Review A}} \textbf{81}, 062343 (2010).

\bibitem{Usenko2016}
V.~C. Usenko and R.~Filip, \enquote{{Trusted noise in continuous-variable
  quantum key distribution: A threat and a defense},}
  {\protect\JournalTitle{Entropy}} \textbf{18} (2016).

\end{thebibliography}

\end{document}